\documentclass[twocolumn,showpacs,preprintnumbers,amsmath,amssymb,superscriptaddress,aps,prc]{revtex4-1}
\usepackage{graphicx}
\usepackage{xcolor}
\usepackage{subfigure}
\def\scr{\rm\scriptscriptstyle }

\begin{document}

\title{Disappearance of Mott oscillations in sub-barrier elastic scattering of identical heavy ions and the nuclear interaction}
\author{L. F. Canto}
\affiliation{Instituto de F\'{\i}sica, Universidade Federal do Rio de Janeiro, CP 68528,
Rio de Janeiro, Brazil}

\author{M. S. Hussein}
\affiliation{Instituto de Estudos Avan\c{c}ados, Universidade de S\~{a}o Paulo C. P.
72012, 05508-970 S\~{a}o Paulo-SP, Brazil, and Instituto de F\'{\i}sica,
Universidade de S\~{a}o Paulo, C. P. 66318, 05314-970 S\~{a}o Paulo,
Brazil}

\author{W. Mittig}
\affiliation{National Superconducting Cyclotron Laboratory and Department of Physics and Astronomy, Michigan 
State University, East Lansing, Michigan 48824, USA}

\begin{abstract}
We investigate the possible disappearance of Mott oscillations in the scattering of bosonic nuclei at sub-barrier energies. 
This effect is universal and happens at a critical value of the Sommerfeld parameter. It is also found that the inclusion of 
the short-range nuclear interaction has a profound influence on this phenomenon. Thus  we suggest that the study of this 
lack of Mott oscillation, which we call, ``transverse isotropy" is a potentially useful mean to study the nuclear interaction.

\end{abstract}

\maketitle

\section{Introduction}
Deviations from pure Mott scattering in the case of heavy-ion systems, has been the subject of investigation over a long period. In particular,  It has been suggested \cite{FMG77, PiG81} that heavy-ion systems, such as $^{12}$C on $^{12}$C and $^{16}$O on $^{16}$O at sub-barrier energies, may exhibit deviation from a pure Mott scattering of identical bosons, owing to the underlying Fermi nature of the constituent nucleons. These authors, invoke the idea of parastatistics to quantify their suggestion. The concept of parastatistics is advanced to describe systems obeying neither the Bose, nor the Fermi statistics, but somewhere in between. Using a parameter that interpolates between the two major statistics, one can test possible deviations of composite bosonic systems such as even-even nuclei from the Bose statistics. Experiments at Yale of $^{12}$C on $^{12}$C and $^{16}$O on $^{16}$O elastic scattering at deep sub-barrier energies
seem to exhibit such deviation, albeit small \cite{Brom77}.

The quest for information about the short range nuclear interaction from elastic scattering data has been going on for a long time. This is even more challenging in the case of elastic scattering of heavy ions, where the long range Coulomb interaction is very important, especially at low energies. This fact prompted researchers to measure the cross section at higher energies, where the Coulomb effects are concentrated in the very small angular region around $\theta = 0$. Useful information was obtained about the nuclear interaction at these higher energies, especially in systems where nuclear rainbow dominates \cite{KOB07, BrS97, HuM84}. One may still wonder if low energy scattering could be used to obtain such information. In fact, it has been shown that information about several useful nuclear properties can be obtained when the energy is below the Coulomb barrier and the cross section is predominantly Coulomb \cite{LTB82, RKL82,HFF84}. Further, any deviation from the Coulomb interaction, even if very small, may lead to measurable change in the characteristics of the Mott oscillations in the scattering of identical nuclei. This fact leads, among other things, to a test of the existence of color Van der Walls force in the Mott scattering of $^{208}$Pb + $^{208}$Pb~\cite{HLP90, VML93}. In this paper we propose to study a special feature of the Mott scattering to obtain information about the short range nuclear interaction, which would, in principle, complement  the information obtained at high energies. This special feature is the apparent disappearance  of the oscillations at a certain critical value of the Sommerfeld parameter. Preliminary work on this has been done in \cite{CDH01}, where the effect was coined Transverse Isotropy (TI). A recent experiment \cite{ASR06} on $\alpha + \alpha$ Mott scattering seems to show this TI. 
Here we go further, and demonstrate and this "Transverse Isotropy", is quite sensitive to the presence of the nuclear interaction, making it an attractive venue to look for the latter.

\section{Identical Particle Scattering}

In the scattering of identical bosons the wave function is symmetric with respect to the 
exchange of the projectile and the target. In the simple case of spin 0, or collisions of 
polarized particles with spin aligned, the spacial part of the wave function must be 
symmetric. The angular distribution must then be given by the expression
\begin{equation}
\sigma(\theta) = \big| f(\theta)+f\left( 90^{\rm o}-\theta \right)  \big|^2,
\label{symmetrized-ftheta}
\end{equation}
where we use the short-hand notation: $\sigma(\theta)\equiv d\sigma/d\Omega$. 
Eq.~(\ref{symmetrized-ftheta}) can be written in the form,
\begin{equation}
\sigma(\theta) = \sigma_{\rm inc}(\theta) + \Xi_{\rm int}(\theta),
\label{inc-int}
\end{equation}
where $\sigma_{\rm inc}(\theta)$ is the incoherent sum of contributions from the two amplitudes,
\begin{equation}
\sigma_{\rm inc}(\theta) = \big| f(\theta)\big|^2+\big|f\left( 90^{\rm o}-\theta \right)  \big|^2,
\label{sig_inc}
\end{equation}
and $\Xi_{\rm int}(\theta)$ is the interference term,
\begin{equation}
\Xi_{\rm int}(\theta) = 2\,{\rm Re}\Big\{ f^\ast (\theta)\,\times f \left(90^{\rm o}-\theta \right) \Big\}.
\end{equation}
Note that the incoherent part of the cross section is positive-definite, whereas the interference term
may assume positive or negative values.\\

Since the scattering amplitude must have a continuous derivative with respect to
$\theta$ and the cross section is symmetric with respect to $\theta = 90^{\rm o}$,
$\sigma(\theta)$ must have a vanishing slope at this angle. 
This poses a few interesting questions like: 
\begin{enumerate}

\item What are the conditions for the angular distribution to have a maximum or
a minimum at $\theta = 90^{\rm o}$? 

\item Can a system present maxima and  minima for different collision energies?

\item If the answer to the previous question is `yes', how does the cross section behaves near
the transition energy?

\end{enumerate}

The aim of the present paper is to answer these questions.

\section{Mott scattering}

We begin by considering a simple scattering problem: the collision of structureless
particles, interacting only through point-charge Coulomb forces. In this case, the
problem has an analytical solution (see, e.g. Ref.~\cite{CaH13}),
\begin{equation}
f_{\rm \scriptscriptstyle{C}}(\theta)=-\frac{a}{2}\  e^{2i\sigma_{0}}\ \
\frac{e^{-i\eta\ln(\sin^{2}\theta/2)}}{\sin^{2}(\theta/2)} ,
\label{fc}
\end{equation}
where $\eta$ and $a$ are respectively the Sommerfeld parameter and half the distance
of closest approach in a head-on collision, given by
\begin{equation}
\eta = \frac{q^2}{\hbar v},\qquad a=\frac{q^2}{2E}.
\end{equation}
Above, $q$ is the charge of the identical particles (the projectile and the target),
$v$ is the relative velocity, and $\sigma_{0}$ is the s-wave Coulomb phase shift
\begin{equation}
\sigma_{0}=\arg \big\{ \Gamma(1+i\eta) \big\},
\label{sig0}
\end{equation}
\smallskip

\noindent with $\Gamma$ standing for the usual Gamma-function.

\medskip

Using these results and normalizing all functions with respect to the Mott cross section at $\theta = 90^{\rm o}$, 
\begin{eqnarray*}
\sigma_{\scr M}(\theta) & \rightarrow&  \overline{\sigma}_{\scr M}(\theta)=\frac{\sigma_{\scr M}(\theta) } 
{\sigma_{\scr M}\left(  90^{\rm o} \right)}\\
\sigma_{\scr inc} (\theta) & \rightarrow&  \overline{\sigma}_{\scr inc}(\theta)=\frac{\sigma_{\scr inc}(\theta) } {\sigma_{\scr M}\left(  90^{\rm o} \right)}\\
\Xi_{\scr int} (\theta) & \rightarrow&  \overline{\Xi}_{\scr int}( \theta )=\frac{\Xi_{\scr int}(\theta) } 
{\sigma_{\scr M}\left(  90^{\rm o} \right)},
\end{eqnarray*}
we can write,
\begin{equation}
{\overline \sigma}_{\scr M}(\theta) = {\overline \sigma}_{\rm inc}(\theta) + {\overline \Xi}_{\rm int}(\theta),
\label{inc-int_bar}
\end{equation}
with
\begin{equation}
\overline{\sigma}_{\rm inc}(\theta )=\frac{1}{16}\,\left[ \frac{1}{\sin ^{4}\left( \theta /2\right) }+\frac{1}{%
\cos ^{4}\left( \theta /2\right) }\right]   \label{siginc/R}
\end{equation}
and
\begin{equation*}
\overline{\Xi}_{\rm int}(\theta ) =\frac{1}{16}\,\left[ 2\ \frac{\cos \left[ 2\eta \ln \left( \tan (\theta
/2)\right) \right] }{\sin ^{2}(\theta /2)\,\cos ^{2}(\theta /2)}\right] .
\end{equation*}
Note that here the cross sections are normalized with respect to the Mott cross section at $\theta = 90^{\rm o}$, whereas in 
Ref.~\cite{CDH01} the normalization was with respect to the Rutherford cross section. These normalizations differ by a factor
4.

\medskip

\begin{figure}[ptb]
\centering
\includegraphics[width=8 cm]{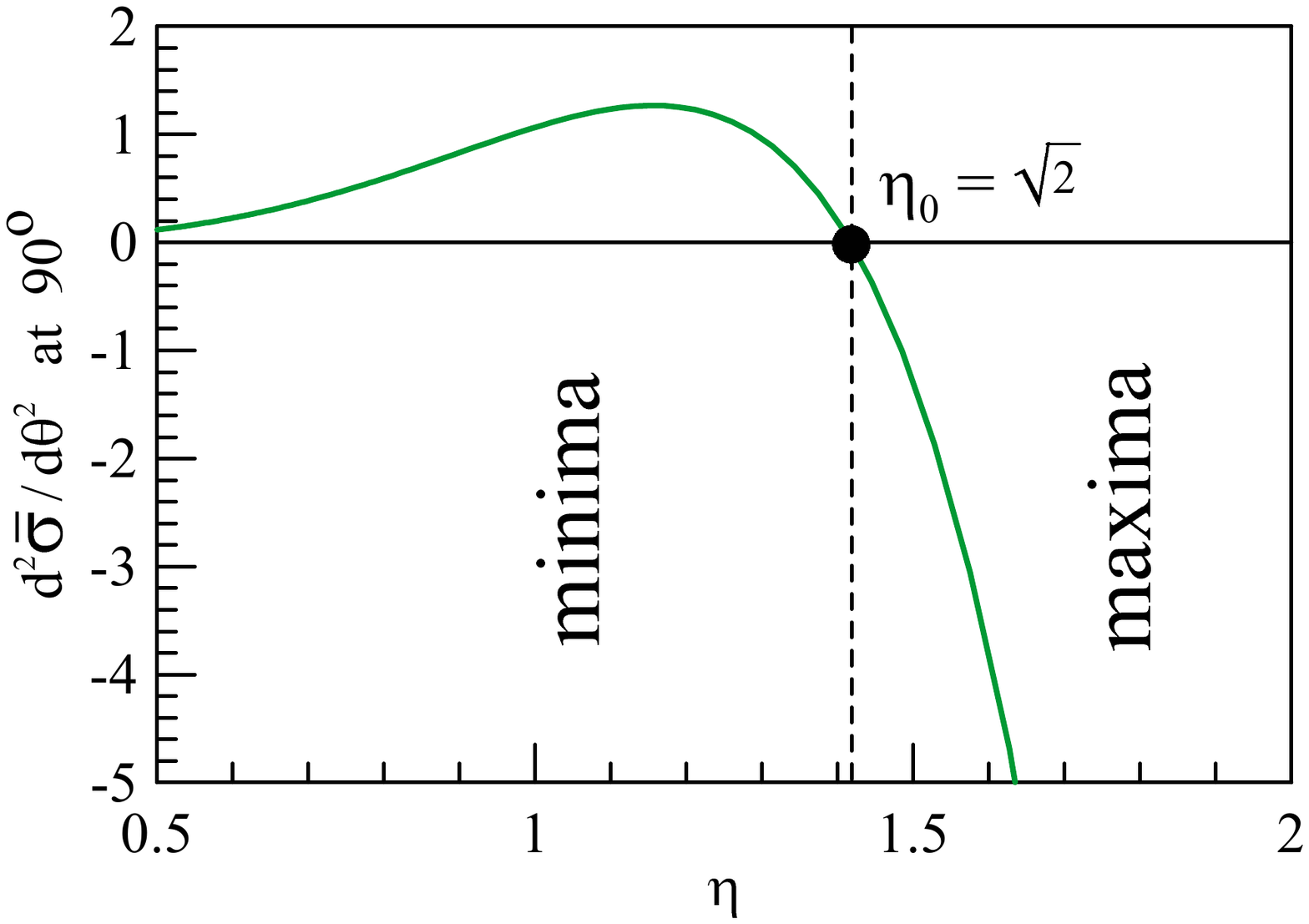}
\caption{(Color online) Second derivative of the renormalised Mott cross section at $\theta = 90^{\rm o}$.}
\label{der2C}
\end{figure}
The sign of the second derivative of the angular distribution at $\theta = 90^{\rm o}$, would indicate wether the cross section has a maximum or a minimum at this angle. Note that the normalized cross section
depends exclusively on the value of the Sommerfeld parameter. In Fig.~\ref{der2C} we show the second derivative
of the angular distribution at $\theta = 90^{\rm o}$, as a function of the Sommerfeld parameter. We see that the
second derivative is positive for small values of $\eta$ and is negative at large values. This means that the angular
distribution has a minima for $\eta<\eta_0$ and a maximum above $\eta_0$. The transition value of the Sommerfeld
parameter can be obtained analytically. After a lengthy calculation~\cite{CDH01}, one obtains $\eta_0=\sqrt{2}$.
\begin{figure}[ptb]
\centering
\includegraphics[width=8 cm]{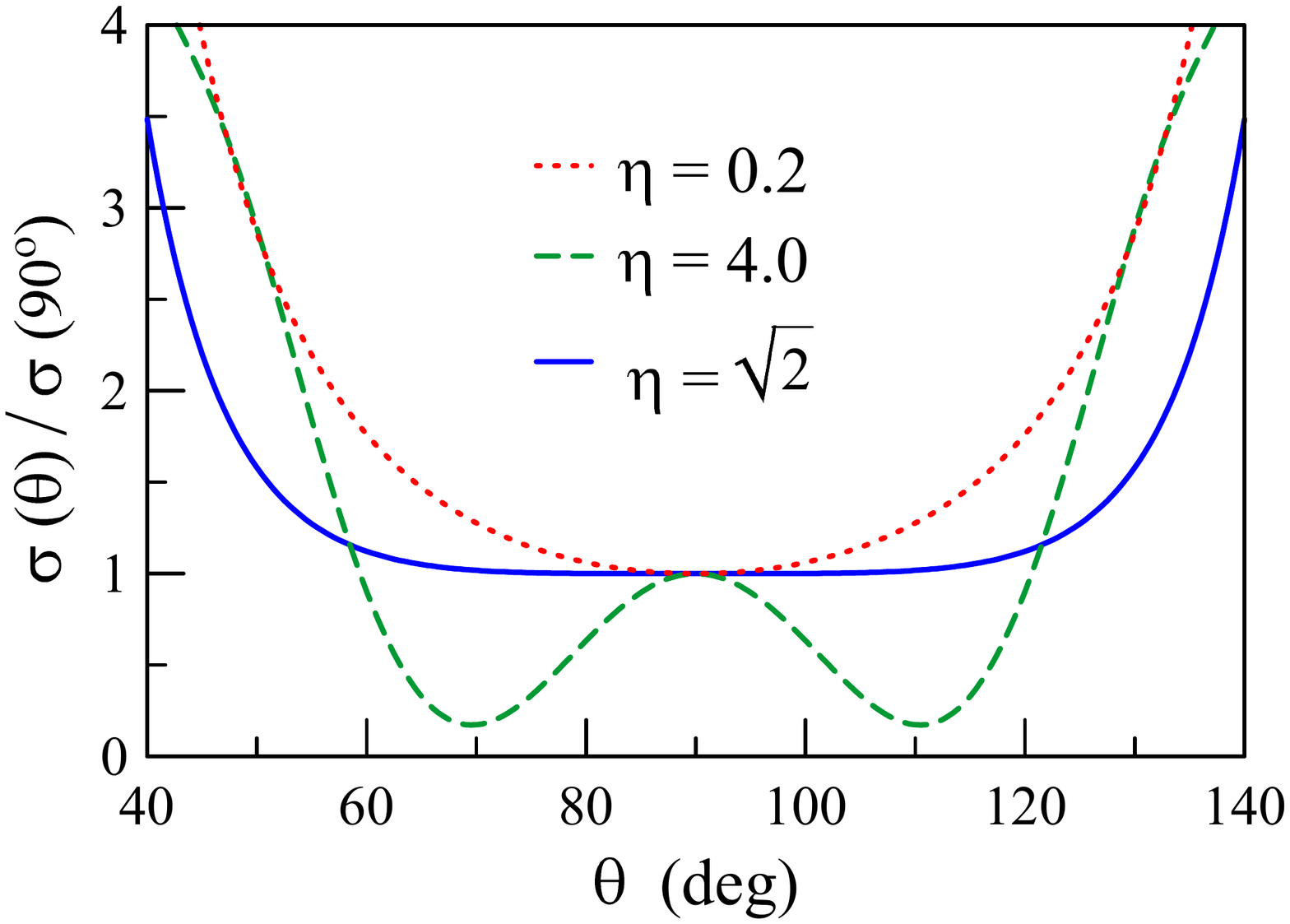}
\caption{(Color online) Mott cross sections for 3 values of the Sommerfeld parameter,
 normalized with respect to their value at $\theta = 90^{\rm o}$.}
\label{sigmott-vs-eta}
\end{figure}

\medskip

To illustrate this behavior we show in Fig.~\ref{sigmott-vs-eta} the normalized Mott cross sections for a value
of the Sommerfeld parameter below $\eta_0$ ($\eta = 0.2$) and one value above ($\eta = 4.0$). As 
expected the former has a minimum at $\theta = 90^{\rm o}$, whereas the latter has a maximum. However, the 
most interesting feature of this figure is the cross section at the critical value of the Sommerfeld parameter,
$\eta =\sqrt{2}$. In this case, the cross section is remarkably flat around $90^{\rm o}$. This phenomenon was
called {\it transverse isotropy} in Ref.~\cite{CDH01}. The important question at this stage is: can this behavior be
observed in some physical system?

\section{Transverse isotropy in Nuclear Physics}

In principle, nuclear systems can be good candidates to exhibit the flat cross sections discussed in the previous
section. However, nuclei interact through both Coulomb and nuclear forces. Thus, the prediction of a flat cross 
section at the critical value of the Sommerfeld parameter will only be valid if the corresponding collision energy
is below the height of the Coulomb barrier. We should then check if there are nuclear systems satisfying this
condition. As qualitative approach to the problem, we set $Z_{\scr P}=Z_{\scr T}=Z$ and
$A_{\scr P}=A_{\scr T}=2Z$, evaluate the collision energy corresponding to $\eta_0$, 
\begin{equation*}
E_0 =\frac{Z^2 e^2}{\hbar \eta_0}\,\sqrt{\frac{m_0\,Z}{2}},
\end{equation*}
and estimate the barrier height by the approximate expression,
\begin{equation*}
V_{\scr B} \simeq \frac{Z_{\scr P}  Z_{\scr T} e^2}{r_0\left( A_{\scr P}^{1/3}+A_{\scr T}^{1/3} \right)}
=\frac{e^2}{r_0}\ Z^{5/3}\,2^{-4/3}.
\end{equation*}
\begin{figure}[th]
\centering
\includegraphics[width=8 cm]{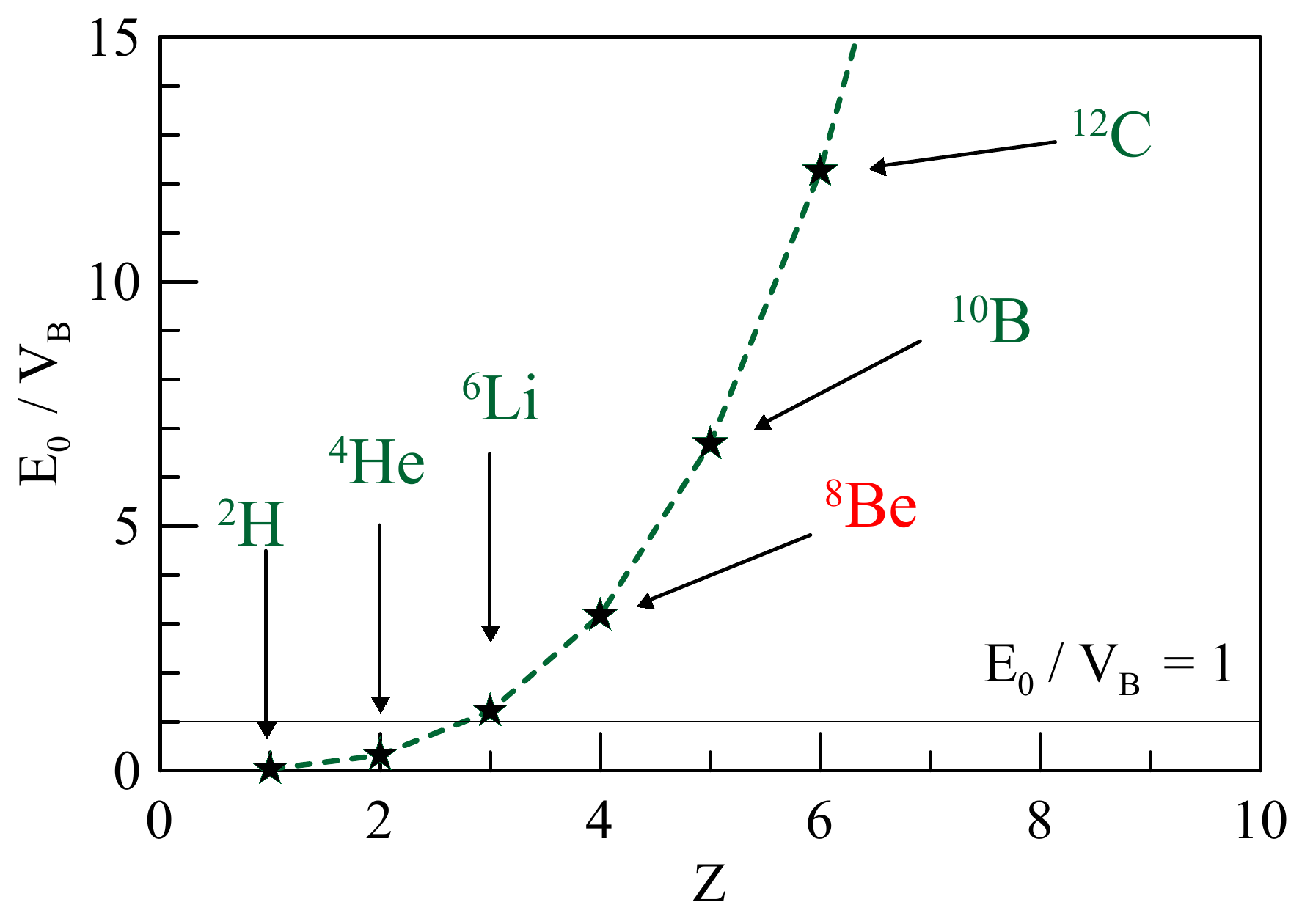}
\caption{(Color online) The ratio of the energy $E_0$ to the Coulomb barrier, $V_{\scr B}$, as a function of the
atomic number of the identical collision partners. For details see the text.}
\label{E0-over-VB-vs-eta}
\end{figure}

In Fig.~\ref{E0-over-VB-vs-eta} we plot the ratio $E_0/V_{\scr B}$ against the atomic number. We see that
only two systems satisfy the condition that the transition energy lies below the Coulomb Barrier: 
$^2$H + $^2$H and
$^4$He + $^4$He. For a third system, $^6$Li + $^6$Li, this condition is nearly satisfied and for heavier
systems the transition energy lies well above the barrier. In the cases of $^2$H + $^2$H and $^6$Li + $^6$Li
the problems is more complicated because these nuclei do not have spin zero. In this way, the above discussion
of the cross section only applies for polarized projectile and target with spin aligned. 
Therefore, we concentrate the discussion to the $^4$He+$^4$He collision.\\

\begin{figure}[th]
\centering
\includegraphics[width=9 cm]{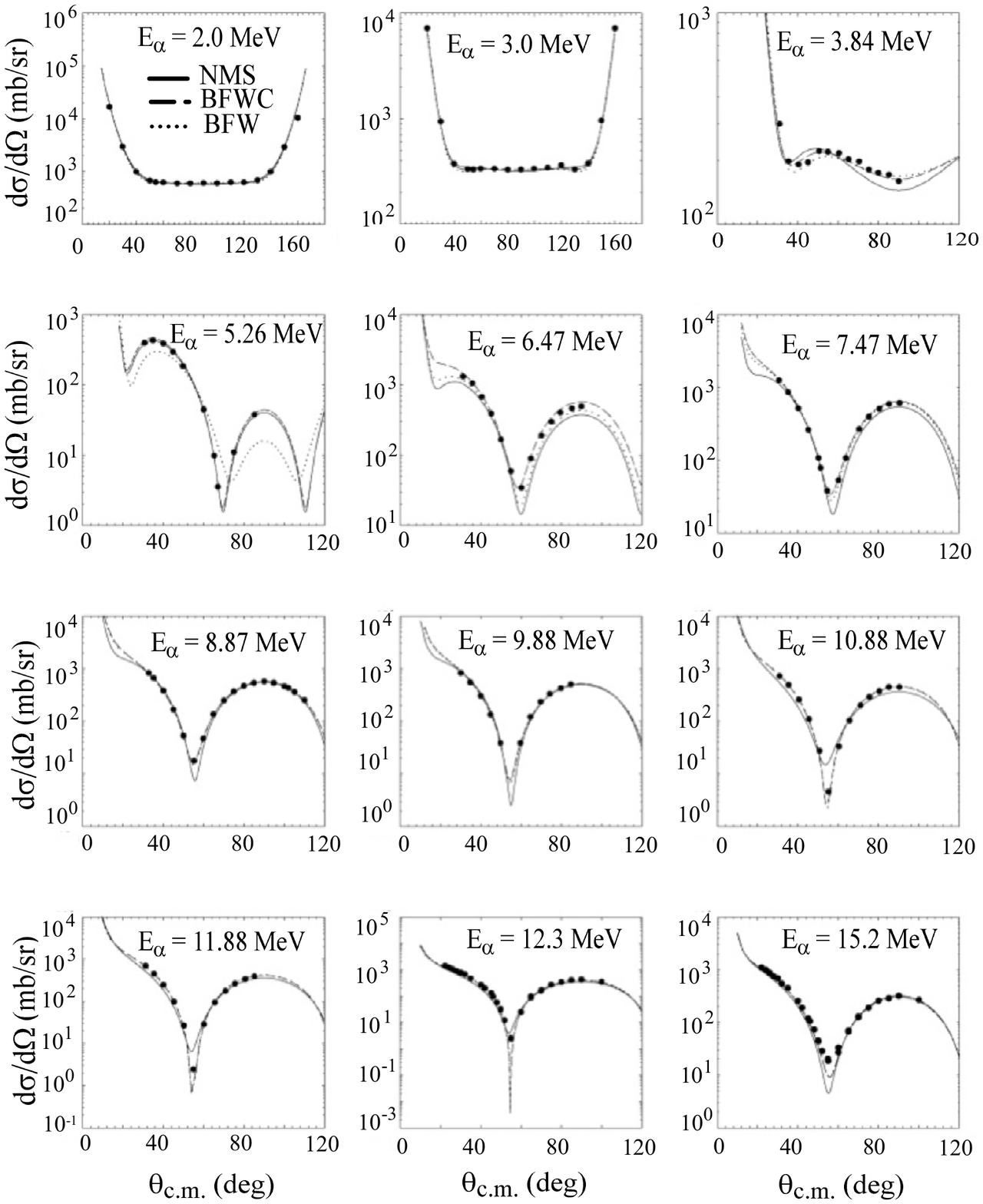}
\caption{(Color online) The experimental angular distributions for the $^4$He+$^4$He system at several
 collision energies. The figure was taken from Abdullah {\it et al.}~\cite{ASR06}. For details see the text.}
\label{abdullah}
\end{figure}

 Fig.~\ref{abdullah} shows experimental angular distributions for the $^4$He+$^4$He system at several
 collision energies. The figure was taken from Abdullah {\it et al.}~\cite{ASR06} (note that the collision energies are
 given in the laboratory frame).  
The lowest energy where data were taken in \cite{ASR06} is $E_{\rm lab}=2.0$ MeV ($E_{\rm c.m.}=1.0$ MeV), which is higher than the collision
energy where the cross section is expected to be flat. For this system, $\eta = \sqrt{2}$ corresponds to 
$E_{\rm c.m.}=0.397$ MeV. However, the data show the flat behavior for the two lowest energies, which
correspond to $E_{\rm c.m.}=1.0$ MeV and $E_{\rm c.m.}=1.5$ MeV. This is surprising and one can only
assume that the nuclear interaction is producing the flat behavior at higher energies. To check this point we
include a nuclear potential in the Hamiltonian and re-evaluate the second derivative of the cross section.
We first try a nuclear potential frequently used to describe nuclear collisions. We consider the Ak\"yuz-Winther
potential~\cite{AkW81,CaH13}. This potential is an approximation to the double folding interaction,  
parametrized by the Woods-Saxon shape,
\[
V(r)= \frac{V_0}{1+\exp\left[ \left(r-R_0\right)/a  \right]}.
\]
The parameters $V_0$, $R_0$ and $a$ are functions of the mass numbers of the projectile and the target 
and for the $^4$He+$^4$He system they have the values: $V_0=-22.21$ MeV, $R_0=3.63$ fm and 
$a=0.5152$ fm. Since the $^4$He nuclei do not have excited states at low energies, we do not include an 
imaginary part in the nuclear interaction. \\

\begin{figure}[th]
\centering
\includegraphics[width=7 cm]{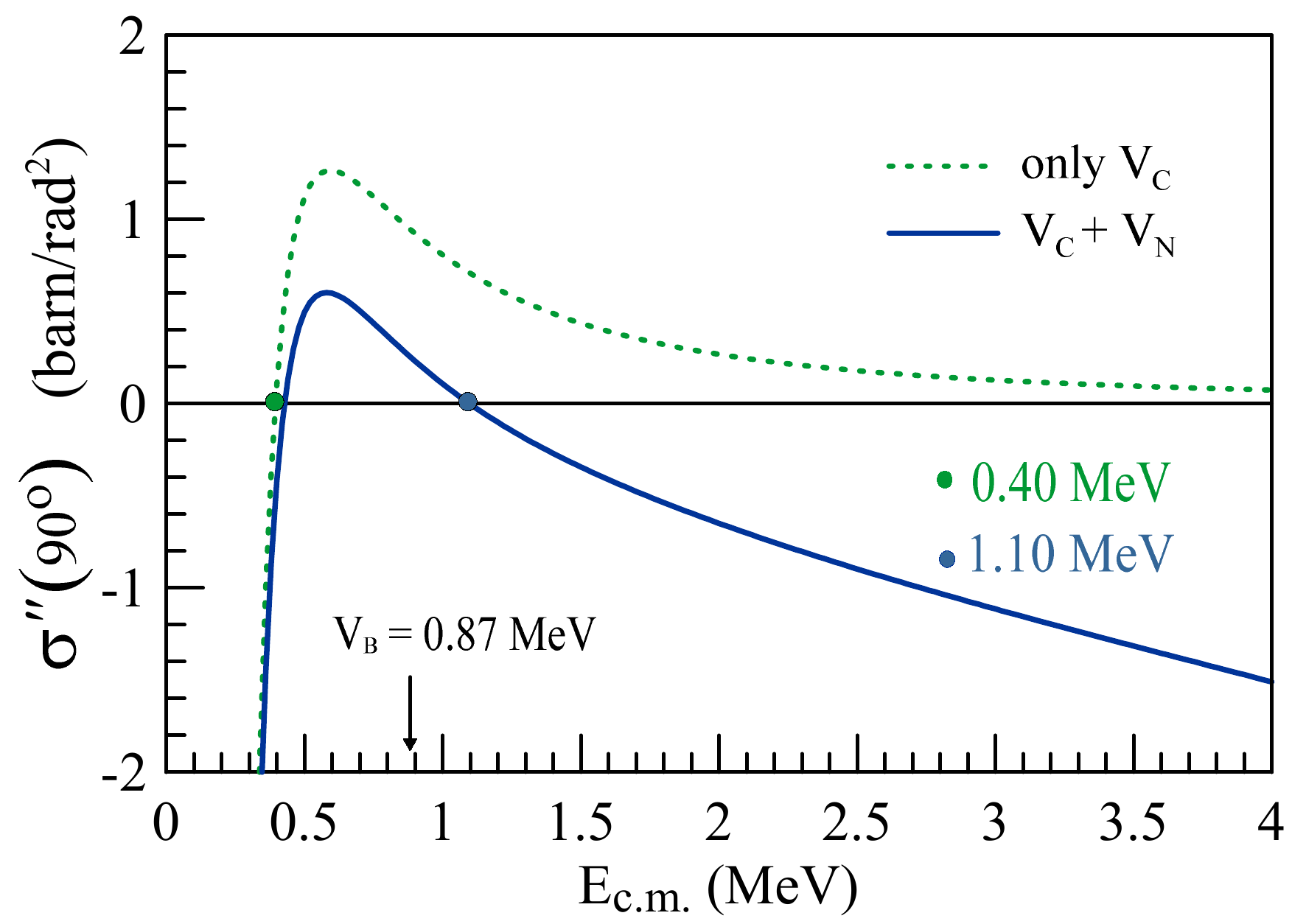}
\caption{(Color online) Second derivative of the Mott cross section at $\theta = 90^{\rm o}$ for the Aky\"us-Winther potential. The dotted line
is basically the same line of Fig.~\ref{der2C}, except that here the plot is against the collision energy. The solid line takes
into account both the Coulomb and the Nuclear potentials. }
\label{dersecC-N}
\end{figure}
In Fig.~\ref{dersecC-N} we show the second derivative of the cross section,
\[
\sigma^{\prime\prime} (\theta)\equiv \frac{d^2\sigma (\theta)}{d\theta^2},
\] 
at $\theta=90^{\rm o}$, both including 
(solid line) and not including (dotted line) nuclear forces. Using the Ak\"yuz-Winther potential, the Coulomb barrier 
for the $^4$He+$^4$He system takes the value $V_{\scr B} = 0.87$ MeV, as indicated within the figure. 
We see that the two curves are very close below $V_{\scr B}$ and the transition energy $E_0$ is not significantly
changed by the nuclear interaction. However, at higher energies the two curves become progressively different.
A very interesting effect of the nuclear force is that it leads to a second transition energy $E_0^\prime = 1.10$ MeV.
In this way, we can distinguish three energy regimes:
\begin{eqnarray*}
{\rm reg.\ 1:}\ \ \ \ \ \ \ \ \, \, \, E< E_0 \ \ &\rightarrow&\ \  \sigma^{\prime\prime}\left( 90^{\rm o}\right)  < 0\ \ ({\rm maximum}) \\
{\rm reg.\ 2:}\ \ E_0<E<E_0^\prime \ \ &\rightarrow& \ \ \sigma^{\prime\prime}\left( 90^{\rm o}\right)  > 0\ \ ({\rm minimum})\\
{\rm reg.\ 3:}\ \ \ \ \ \ \  \, \, \, E_0>E_0^\prime \ \ &\rightarrow&\ \  \sigma^{\prime\prime}\left( 90^{\rm o}\right) < 0\ \ ({\rm maximum}).
\end{eqnarray*}

\noindent In regions 1 and 3 the second derivative is negative, hence $\sigma(90^{\rm o})$ is a maximum. On the other hand,
in region 2 the second derivative is positive and $\sigma(90^{\rm o})$ is a minimum.\\

Fig.~\ref{dersecC-N} explains some features of the experimental cross sections of Fig.~\ref{abdullah}. First,
it predicts a second region where the cross sections is flat. This regions is around $E_{\rm c.m.}\sim 1$ MeV,
as in the data of Abdullah {\it et al.}~\cite{ASR06}.  Above this region, Fig.~\ref{dersecC-N} predicts that
cross section will have a maximum at $\theta=90^{\rm o}$. This is correct, with one exception: the data
at $E_{\rm lab}=3.84$ MeV. At this energy the experimental cross section has a minimum, whereas following Fig~\ref{dersecGauss}, a maximum is predicted.This point should be studied more carefully.\\

\begin{figure}[th]
\centering
\includegraphics[width=8 cm]{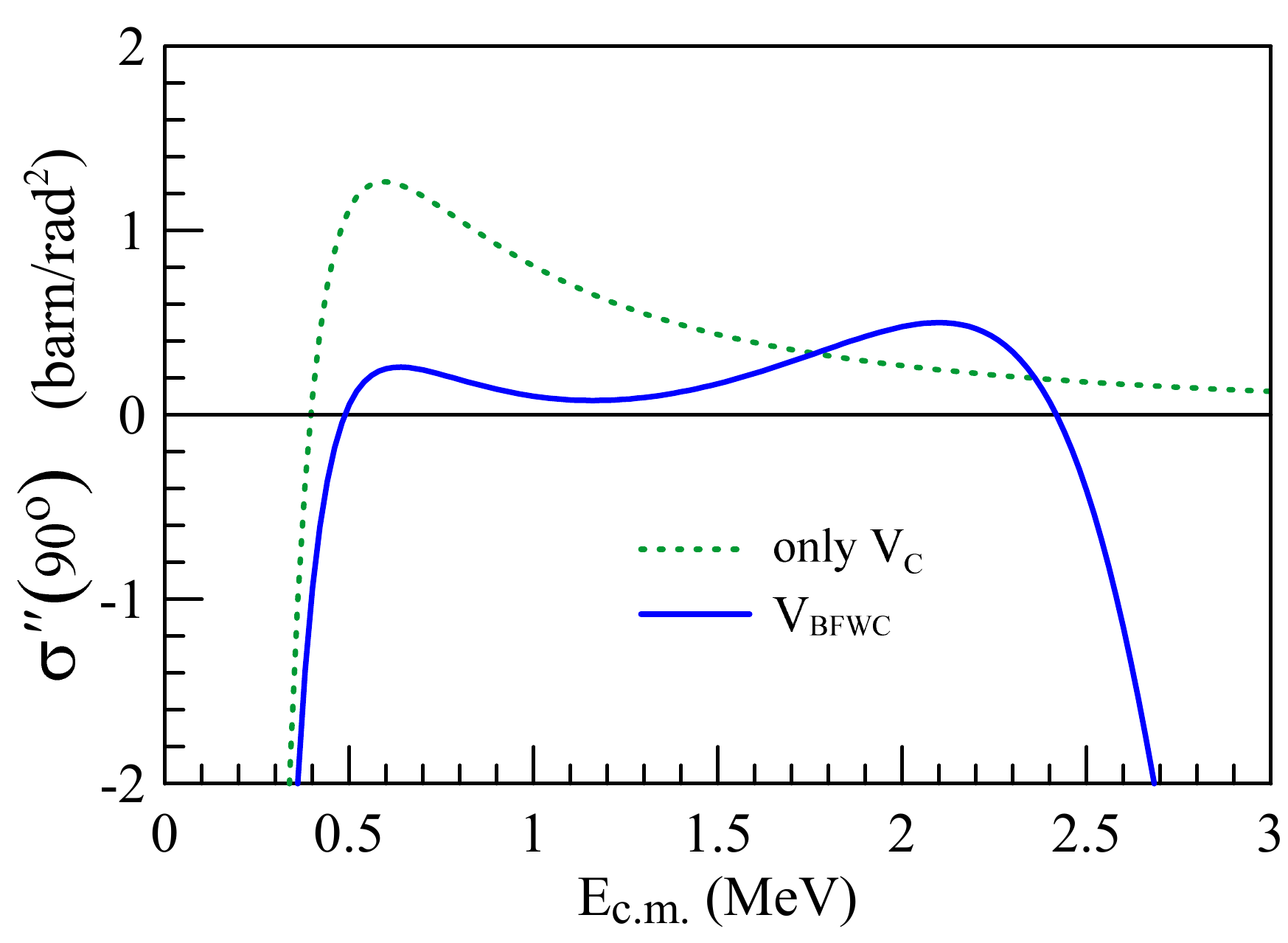}
\caption{(Color online) Same as Fig.~\ref{dersecC-N} but now the nuclear potential is given by
Eqs.(10-12), with the parameters given in the text.}
\label{dersecGauss}
\end{figure}
The Ak\"yuz-Winther interaction is quite successful  for the description of heavy ion collision. However, it
is not suitable for $^4$He. The data in the $\alpha$ energy region $E_{\alpha} $= 2 - 34.2 MeV, were 
accounted for by a real nuclear interaction given by the sum of two gaussians, one attractive and one repulsive,
plus the Coulomb interaction. This potential, which in the work of Abdullah {\it et al.} \cite{ASR06} was called 
BFWC, is given by,
 \begin{equation}
V_{\scr BFWC}(r)  = - \,V_{\scr A}\ e^{- r^2/R^2_{\scr A}} + V_{\scr R}\  e^{- r^2/R^2_{\scr R}}+V_{\scr C}(r).
 \end{equation}
Above, $V_{\scr C}(r)$ is the Coulomb interaction,
\begin{eqnarray}
V_{\scr C}(r)   &=& \frac{4e^2}{2R_{\scr C}}\left( 3 - \frac{r^2}{R^2_{\scr C}}\right),\qquad {\rm for}\ r< R_{\scr C}\\
 &=& \frac{4e^2}{r}\qquad\qquad\qquad\qquad {\rm for}\ r \ge  R_{\scr C},
\end{eqnarray}
with $R_{\scr C} = 5. 8$ fm, and the parameters of the nuclear potential are $V_{\scr A}$ = 122.62 MeV, $R_{\scr A}$ = 2.132 fm, 
$V_{\scr R}$ = 3.0 MeV and $R_{\scr R}$ = 2.0 fm.
 
\smallskip

We then adopt this potential and evaluate the second derivative of the
cross section at $\theta = 90^{\rm o}$, as a function of the collision energy. The result is given in 
Fig.~\ref{dersecGauss}. The lowest transition energy shifts slightly, taking
the value $E_0=0.47$ MeV and the second one moves to $E_0^\prime = 2.41$ MeV. Between these two 
energies the second derivative remains very small, getting very close to zero around 2.2 MeV. The energy
$E_{\rm lab}=3.84$, where the experimental cross section shows a minimum at 90$^{\rm o}$ is in the region
where the second derivative grows before dropping as the energy approaches $E_0^\prime$.\\

The results shown in Fig.~\ref{dersecGauss} imlpy that the cross sections must be very flat at the
energies $E_{\rm c.m.}  = 0.47$ MeV and $E_{\rm c.m.}  = 2.41$ MeV and between these values
the cross section has a slight minimum. Above  2.41 MeV, the cross sections
present pronounced maxima at 90$^{\rm o}$. This is effectively the experimental behavior \cite{ASR06}. No experimental results are available for the other predicted transition point at
$E_{\rm c.m.} = 0.47$ MeV.

\section{Conclusions}

We have discussed the effect of the nuclear interaction on the Transverse Isotropy, namely, the angular region where the Mott 
cross section becomes flat. Application was made for the $\alpha + \alpha$ system, where data exists at the near-barrier
 energies considered \cite{ASR06}. We have found an important sensitivity to the nuclear interaction. Our finding should 
 be helpful to investigate the nuclear interaction in the Mott scattering of heavy ions. In particular the transition region with Transverse Isotropy predicted at $E_{\rm c.m.} = 0.47$ MeV in the $^{4}$He + $^{4}$He system should give a precise determination of the long range part of the nuclear potential.\\

Acknowledgments\\

Partial support from the CNPq and FAPESP are acknowledged.

\bibliographystyle{apsrev}    
\bibliography{mott_01}   

\end{document}